\documentstyle [preprint,eqsecnum,aps]{revtex}

\begin{document}
\draft
\preprint{SNUTP-93-96}

\title{Abelian Chern-Simons field theory \\
and anyon equation on a torus}

\author { Kyung-Hyun Cho and Chaiho Rim}

\address
{Department of Physics\\
Chonbuk National University\\
Chonju, 560-756,Korea}

\date {December 24, 1993}

\maketitle

\begin{abstract}

We quantize the abelian Chern-Simons
theory  coupled to non-relativistic
matter field  on a torus without invoking the flux
quantization.
Through a series of canonical transformations
which is equivalent to solving the Gauss
constraint,  we obtain an effective hamiltonian density
with periodic matter field.
We also obtain the many-anyon Schr\"odinger equation
with periodic Aharonov-Bohm potentials
and analyze the periodic property of the
wavefunction.
Some comments are given on
the different  features of our approach from the previous ones.

\end{abstract}


\narrowtext

\section{Introduction}

Chern-Simons theory \onlinecite{deser,hagen}  in 1+2
dimensions has attracted much interest
because it exhibits various theoretically
interesting aspects.
The particles of fractional statistics (anyons) \cite{leinaas}
are described as particles of ordinary statistics
interacting  with a Chern-Simons
gauge field \onlinecite{goldhaber,semenoff,matsuyama,swanson,cho}.
It presents an effective  description of the statistical
transmutation.  When the theory is
coupled to a certain scalar field theory,
the system possesses stable topological
and non-topological vortex solutions \cite{hong}
whose interpretation \cite{kim} as anyons is also suggested.

Quantum mechanics of many
anyons is actively investigated
last few years in relation with its direct
relevance to the fractional  quantum Hall effect
and the related hierarchical
models and its speculative  role
in high $T_c$ super-conductivity \cite{girvin}.
The construction of many anyon wavefunction
is not easy since even its free
form can not be written as a product
of free single-body wavefunctions, although
there are some analytical methods \cite{wu}
found for simple models on an infinite
plane. In addition, the analysis of many anyons
is plagued with the
hard-core boundary condition and the idea of
smooth statistical  interpolation of
the Hilbert space from  boson to fermion
(or fermion to boson)  waits further
clarification \cite{amelino} whose numerical
try \cite{sporre} has been done for a few-body system.

Anyons on a compact space need  extra-care
since they require consistent boundary
conditions at the edge or some invariance property
on a closed surface. In
addition, zero-modes of the gauge field
complicate the analysis. It is well
known \cite{polychro} that the vacuum  on a torus is
degenerate due to the zero-modes
of the gauge fields  and the degeneracy
is finite if the  Chern-Simons
coefficient is rational.

Many particle quantum mechanics on the torus  is also
constructed \onlinecite{randjbar,iengo,hosotani,fayya}
from the abelian Chern-Simons
field theory like in the infinite
plane case \cite{jackiw,kimlee}, and it is demonstrated
that the degeneracy structure of the
vacuum can affect the quantum mechanics
and give rise to a multi-component
anyon wavefunction \onlinecite{hosotani,fayya,ein}.
It seems, however, that at least two essential
points,  quantization and periodic property
of the matter field,  are to be
clarified.  The problem related
with the quantization can arise  when one
solves the Gauss constraint before quantization.
It is noted \cite{hagsu} that
treating the number operator
as a $c$-number by restricting the Hilbert space
on the fixed number of particles  may lead to subtleties.
Closely related with
this is the quasi-periodic  boundary condition on the fields
and the statistical flux quantization.
To avoid the possible subtleties,  it seems
desirable to quantize
all the fields before solving the Gauss
constraint  and analyze the translation
invariance of the system  on the torus
to discover the periodic property of the fields.

In this paper, we re-analyze the Chern-Simons theory
coupled to  non-relativistic matter fields
on a torus  by following the procedure done in Refs. \cite{swanson,cho}
for the infinite plane
case. Namely, we quantize the theory first  and through canonical
transformations
we obtain an effective theory
of matter fields only by removing the gauge fields including the
zero-modes,  which is equivalent
to solving the Gauss constraint. In this
canonical formalism, it turns out
that the statistical flux  quantization is not
necessary. The translation invariance
along the non-contractible loops is
manifest as well as the gauge invariance
irrespective of the  particle numbers.
The transparent periodic property of
fields enables us to deduce that of the
original fields.  From this, we also obtain the periodic
property of  the gauge
invariant wavefunction of many anyons
and construct the  Schr\"odinger equation
for that.

This paper is organized as follows.
In section II, we briefly describe
the quantization of abelian pure Chern-Simons gauge field
on a finite plane (with the
periodic boundary condition) with an eye
to the zero-modes of the gauge field.
In section III, we obtain an effective hamiltonian
of matter field in the
fundamental domain of the torus
by defining  new fields through a series of
canonical transformations.
In section IV, we construct the many-body
Schr\"odinger equation in the fundamental
domain of the torus.  The equation becomes the
free one  except the Aharonov-Bohm potential
which is responsible for the
statistical  transmutation.
In section V, we consider the periodic property of
the system.  We construct the effective
hamiltonian density such that it is
translation invariant along the
non-contractible loops by defining a simple
periodic matter field in consistency
with two commuting translation operators.
The periodic property of the original fields
is also presented through canonical
transformations.  We obtain the
Schr\"odinger equation of  many anyons with
periodic Aharonov-Bohm potential.
In section VI, we summarize our results  and
discuss the different aspects of
our analysis from the  previous ones.

\section{Mode expansion of gauge field}

Lagrangian density of abelian pure Chern-Simons
theory on an infinite plane is given as
\begin{equation}
{\cal L}= {\mu \over 2}
\epsilon^{\mu\nu\rho} a_\mu \partial_\nu a_\rho\,.
\end{equation}
The gauge field does not propagate
and the hamiltonian system becomes a constraint one.
The phase space variables are  $a_1$ and $\pi_1$,
where
\begin{equation}
\pi_1 = {\partial {\cal L} \over \partial \dot a_1}
= {\mu \over c} a_2\,,
\end{equation}
and the gauge-field itself constitutes the phase space.
Its equal-time
commutation relation is given as
\begin{equation}
[a_1(x), a_2(y)] = i{\hbar \over \mu}
\delta (\vec x - \vec y)\,,
\label{a12comm} \end{equation} where $x$
denotes the three vector $(ct, \vec x )$ and
$\vec x =(x^1, x^2) = (-x_1, -x_2)$.
$a_0(x)$ is treated as a lagrange
multiplier and  commutes with $a_i$'s.

The hamiltonian density becomes
(ignoring the space-time boundary terms),
\begin{equation}
{\cal H} = \mu a_0 b \,,
\end{equation}
where $b  = - (\partial_1 a_2 - \partial_2 a_1)$ is
the statistical magnetic field.
This shows that $b$ is the gauge generator.
The physical state $|{\rm phys}>$ should be
annihilated by $b$,
$b |{\rm phys}> = 0$.

To understand the physical state more deeply,
we recall that the gauge operators $a_i$'s
on an infinite plane is expanded as \cite{cho}
\begin{eqnarray}
a_i(x)&=&
\int {d \vec p \over 2 \pi}
{1 \over \sqrt { 2 p_0}}
[ ( {p_i \over \mu} g(\vec p)
+ i{\epsilon_{ij} p_j \over p^0} h (\vec p))
e^{-ip\cdot x} + c.c]_{p^0 = |\vec p|}
\nonumber\\
a_0(x)&=&
\int {d \vec p \over 2 \pi}
{1 \over \sqrt {2 p_0}}
[ {p_0 \over \mu} g(\vec p) e^{-ip\cdot x}
+ c.c]_{p^0 = |\vec p|}\,.
\label{a01def}
\end{eqnarray}
$g(\vec p)$ and $h(\vec q)$ satisfy
the commutation relations,
$$ [g(\vec p), h^+(\vec q)]
= \hbar \delta(\vec p - \vec q)
$$
\begin{equation}
[g(\vec p), h(\vec q)]
=[g(\vec p), g^+(\vec q)]
= [h(\vec  p ), h^+(\vec q)]
=0\,.
\label{ghcomm}
\end{equation}

The statistical magnetic field $b$ is given as
\begin{equation}
b(x) =\int {d \vec p \over 2 \pi}
{1 \over \sqrt {2 p_0}}
[p^0 h(\vec p) e^{-ip\cdot x} + c.c]_{p^0 = |\vec p|}\,.
\end{equation}
This representation is constructed
such that the Lorentz gauge fixing condition
is satisfied  identically,
$\partial_\mu a^\mu = 0$ and  $[ a_0 , a_i]=0$.
We note that there is still left a residual
gauge  degree of freedom with the
Lorentz gauge fixing condition maintained:
$a_\mu \rightarrow a_\mu + \partial_\mu  \Lambda$
where $\partial^2 \Lambda=0$.
This merely redefines $g(\vec p)$ in Eq.
(\ref{a01def}) satisfying the same commutation
relation in Eq. (\ref{ghcomm}).
In our analysis,
we do not need the specific form of $g(\vec p)$ and therefore,
we can proceed in the residual gauge independent manner.

On a finite plane, the gauge field is given
in terms of summation of discrete modes,
which should include non-commuting zero-modes
$\theta_1$ and $\theta_2$ to satisfy the
commutation relation  Eq.(\ref{a12comm}) on the box.
For definiteness, we will consider  the finite plane
(the fundamental domain of the torus) as the
rectangular box with (period) length $L_1$ and $L_2$
in the following. If $b$ is
periodic with a mean magnetic field,
then $a_i$ is given as \begin{eqnarray}
a_i(x) &=&  {\theta_i(t) \over L_i} + \tilde x_i \xi(t)
\nonumber\\
&+& {1 \over \sqrt {L_1L_2}}
\sum_{\vec p \ne 0} {1 \over \sqrt {2p^0}}
[ ( {p_i \over \mu} g(\vec p) + i{\epsilon_{ij} p_j
\over p^0} h (\vec p)) e^{-ip\cdot x} +
c.c]_{p^0 = |\vec p|}\,, \label{a1box}
\end{eqnarray}
where $p_i = {2\pi n_i \over L_i}$ with $i =1,2$.
$\theta_i$'s are two non-commuting real zero-modes;
\begin{equation}
[\theta_1, \theta_2] = i {\hbar \over \mu } \,.
\label{theta12comm}
\end{equation}
$\theta_i$'s commute with the rest of the other modes.
The $\theta_i$ modes are essential
to satisfy the  commutation relation in
Eq. (\ref{a12comm}) on the box.
We introduced the mean magnetic flux mode $\xi$,
which commutes with the other modes and
\begin{equation}
\tilde x_1 \equiv \phi {x^2 \over L_1L_2}\,;\quad
\tilde x_2 \equiv 0\,,
\end{equation}
where $\phi$ is the
mean statistical magnetic flux.
$g$ and $h$ satisfy the discrete version of
the commutation relation, Eq. (\ref{ghcomm}).

On the other hand, $a_0(x)$ does
not contain  the zero-modes
and is expressed in the discrete  form of Eq.
(\ref{a01def}) with an addition of a time independent
constant Lagrangian multiplier $\lambda$
(which is convenient to describe the system  coupled with
matter field),
\begin{equation} a_0(x)
= \lambda  + {1 \over \sqrt {L_1L_2}}
\sum_{\vec p \ne 0} {1 \over \sqrt {2 p_0}}
[ {p_0 \over \mu} g(\vec p) e^{-ip\cdot x}
+ c.c]_{p^0 = |\vec p|}\,.
\label{a0box}
\end{equation}
$a_0$ commutes with other fields.
We note that the Lorentz gauge fixing condition
is satisfied identically
even though the zero-modes,
$\theta_i$ are time-dependent.

Since there are two
non-commuting zero-modes,
which commute with the hamiltonian density
$[\theta_i, {\cal H}] =0$
(recall $b$ and $a_0$ do not contain the zero-modes),
the physical state $|{\rm phys}>$ can be degenerate.
The degeneracy structure is interesting
in itself \onlinecite{polychro,iengo,hosotani,fayya}
but we do not elaborate on this.
As is seen in section III and V,
the gauge sector of the vacuum can be decoupled
from the matter sector
and  one can choose many anyon wavefunction
such that the  details of the vacuum
degeneracy do not enter.
Other choice of the wavefunction and its relation
with the vacuum degeneracy is discussed
in section VI.

\section{Matter-coupled  system}
Let us consider the system in the fundamental
domain of the torus,
$D_{(0,0)}$, ($0 \le x^1 < L_1$
and $ 0\le x^2 <L_2$).
The lagrangian density of matter field and Chern-Simons
gauge field is given as
\begin{equation} {\cal L}= {\mu \over 2}
\epsilon^{\mu\nu\rho}
a_\mu \partial_\nu a_\rho\ + \psi^+ i\hbar D_0 \psi -
{\hbar^2 \over 2m} |\vec D \psi|^2\,,
\end{equation}
where
$i\hbar D_\mu \psi =
(i \hbar \partial_\mu
- {e\over c } a_\mu ) \psi$.
We will choose the matter field
$\psi$ as a fermion-field for definiteness.
The following analysis goes parallel
with bosonic one  for the above lagrangian.
The hamiltonian density is written as
\begin{equation}
{\cal H}=
a_0 (\mu b + {e\over c}J_0)
+ {1\over 2m} |i\hbar D_i \psi|^2\,,
\end{equation}
where $J_0 = \psi^+ \psi$.
The operators satisfy the equal-time
(anti-)commutation  relations;
$$ \{\psi(x), \psi^+(y)\} =  \delta (\vec x - \vec y) $$
$$ [J_0(x), \psi^+(y)] = \psi^+(x) \delta (\vec x - \vec y) $$
\begin{equation}
[a_1(x), a_2(y)] = i {\hbar \over \mu}
\delta(\vec x - \vec y)\,.
\label{commfield}
\end{equation}
The gauge field $a_i$ has the mode
expansion in Eq.(\ref{a1box})
and $a_0$ is given in Eq. (\ref{a0box}).
The constraint is given as the Gauss law,
$\Gamma = 0$ where \begin{equation}
\Gamma = b + {e \over \mu c} J_0
\end{equation}
is the gauge generator.
(Note that due to the $\lambda$ in Eq. (\ref{a0box})
the Gauss law includes the mean field component.)
Therefore, a physical state should be annihilated by $\Gamma$,
$\Gamma |{\rm phys}> = 0$.

A commonly adopted procedure  for this system is to
solve the Gauss constraint  explicitly
before quantizing the fields. On an
infinite plane, this method is equivalent
to the one with the reverse process,
{\it i.e.} quantizing first and requiring constraint
on the Hilbert space.
On a finite plane, however,
solving the constraint  first can lead to subtleties as
mentioned in the Introduction.
To avoid this, we quantize the field first  as
in Eq. (\ref {commfield}) and perform a canonical
transformation of this system to get
a new effective hamiltonian density
in terms of  newly defined fields and the
physical states are explicitly constructed.
To give an overview on our method,
we summarize our procedure as follows.

The hamiltonian density ${\cal H}$ is
the functional of $\psi(x)$, $a_i(x)$, and $a_0(x)$.
$$ {\cal H} = {\cal H}
(\psi(x), a_i(x), a_0(x))\,. $$
We will define new fields with a canonical
transformation operator $V(t)$ which commutes with  $a_0(x)$ as
\begin{eqnarray}
\psi(x) &=& V(t) \psi^{(N)}(x) V^+(t)
\nonumber\\
a_i(x) &=&V(t) a_i^{(N)} (x) V^+(t)\,.
\label{fieldcan}
\end{eqnarray}
It is obvious
that the new fields satisfy the same commutation relations
in Eq. (\ref{commfield}).
Then ${\cal H}$ is given as  canonical transformation of the
hamiltonian density with the new fields
$$ {\cal H}(\psi(x), a_i(x), a_0(x))
= V(t) {\cal H}(\psi^{(N)}(x), a_i^{(N)}(x), a_0(x)) V^+(t)\,. $$
Finally we express the same hamiltonian
as a new functional of the new fields as
\begin{equation}
{\cal H}(\psi(x), a_i(x), a_0(x))
= {\cal H}^{(N)}(\psi^{(N)}(x), a_0(x))\,.
\end{equation}

The merit of this procedure lies in that we can choose
$V(t)$ such that $\psi^{(N)}(x)$ is gauge  invariant
\begin{equation}
[\psi^{(N)}(x), \Gamma(y)] =0\,,
\label {comment}
\end{equation}
and the effective hamiltonian density contains
the gauge invariant fields only.
In fact, we choose the newly defined
statistical magnetic  field as the gauge generator.
Therefore, the Gauss constraint is satisfied automatically
if the vacuum is defined as the (local)
gauge invariant state
(which has neither flux nor particle in the original picture).
The physical states are obtained
if the product of the newly defined matter field
$\psi^{(N)+}$ is applied on the vacuum.

To substantiate the above procedure
clearly, we will take the three steps of canonical transformations,
$V= V_1V_2V_3$.
Let us consider a canonical transformation operator
\begin{equation}
V_1= \exp i \left[ Q \eta^{(1)}
+ {e\over \hbar c} \int\! \int d\vec x d \vec y
J_0^{(1)}(\vec x,t) G_p (\vec x, \vec y)
\partial_ka_k^{(1)}(\vec y, t) \right] \,.
\label{V1}
\end{equation}
$Q \equiv \int d \vec x J_0(x)$ is the number operator.
$G_p$ is the periodic (single-valued on the torus)
Green's function   satisfying
\begin{equation}
\nabla^2 G_p (\vec x, \vec y)
= \delta (\vec x, \vec y)
- {1 \over L_1 L_2}\,; \quad
G_p(\vec x, \vec y) = G_p (\vec y, \vec x)\,,
\end{equation}
and is given by
\begin{equation}
G_p(\vec  x, \vec y) = {1\over 4\pi}
\ln \left|
{\theta_1(z|\tau) \over \theta_1{'}(0|\tau)}
\right|^2  -
{({\rm Im } \, z)^2 \over 2 \,
{\rm Im} \, \tau}\,,
\end{equation}
where $z= [(x^1-y^1) +
i (x^2 - y^2)]/L_
1$, $\tau = i L_2 /L_1$
and $\theta_1(z|\tau)$ is the  odd Jacobi theta
function.
We note that it is regularized such that
\begin{equation}
\epsilon_{ij} \partial_{x^j} G_p(\vec x, \vec y)
\left.\right|_{\vec x = \vec y}  =0\,.
\end{equation}
$\eta^{(1)}$ in Eq. (\ref{V1})
is a conjugate to $\xi^{(1)}$ satisfying
equal-time commutation relation,
\begin{equation}
[\eta^{(1)}, \xi^{(1)}] = i\,.
\end{equation}

The original fields are given  in terms  of the new fields as
\begin{eqnarray}
\psi(x) & =& V_1 (t) \psi^{(1)}(x) V^+_1(t)
\nonumber\\
& =& \psi^{(1)}(x)
\exp -i \{\eta^{(1)} + {e\over \hbar c}
\int \! d \vec y
G_p(\vec x, \vec y)  \partial_{y^i} a_i^{(1)} (y) \}\,,
\nonumber\\
a_i(x) &=&
V_1(t) a_i^{(1)}(x) V_1^+(t)
\nonumber\\
&=& a_i^{(1)}(x) -
Q \tilde x_i - {e \over \mu c}
\epsilon_{ij} \partial_{x^j}
\int \! d\vec y G_p(\vec x, \vec y) J_0^{(1)} (y)\,,
\nonumber\\
\xi(t) &=& V_1(t) \xi^{(1)}(t) V_1^+(t)
= \xi^{(1)}(x) - Q\,,
\nonumber\\
\eta(t) &=& V_1(t) \eta^{(1)}(t) V_1^+(t)
= \eta^{(1)}(t)\,,
\label{field1def}
\end{eqnarray}
$\theta_i(t) = \theta_i^{(1)}(t)$,
and $J_0(x)= J_0^{(1)}(x)$.
To have the relation  $\Gamma (x) = b^{(1)} (x)$,
we choose the flux $\phi = {e\over \mu c}$,
which simplifies the analysis for the constraint system.
We replaced the superscript $(N)$
in Eq. (\ref {fieldcan})  by $(1)$ for an obvious reason.
The hamiltonian density ${\cal H}^{(1)}$
is given in terms of  the new fields,
\begin{equation}
{\cal H}^{(1)} = \mu a_0 b^{(1)}
+ {1 \over 2m}  \left|
(i \hbar \partial_i - {e\over c} a_i^{({\rm eff})} )
\psi^{(1)} \right|^2\,.
\label{hamil1}
\end{equation}
Implicit ordering of $\psi^{(1)+}$ to the left of $a_i^{({\rm eff})}$
is assumed.
Here $a_i^{({\rm eff})}$ is an effective gauge-like term,
\begin{equation}
a_i^{({\rm eff})}(x)
= {\theta_i \over L_i}
- (Q +1) \tilde x_i
+\epsilon_{ij} \partial_{x^j}
\int \! d \vec y G(\vec x, \vec y)
b^{(1)}(y) -{e \over \mu c}
\epsilon_{ij} \partial_{x^j}
\int \! d \vec y G_p(\vec x, \vec y)J_0(y)
\,. \label{aeff}
\end{equation}
This is obtained by noting that (see Eq. (\ref{a1box}))
\begin{equation}
a_i^{(1)}(x)
= {\theta_i \over L_i}
+\epsilon_{ij} \partial_{x^j}
\int \! d \vec y G(\vec x, \vec y) b^{(1)}(x)
+ \partial_{x^i} \int \! d \vec y G_p(\vec x, \vec y)
\partial_{y^j} a_j^{(1)} (y) \,,
\label{a1i}
\end{equation}
where
$$ G(\vec x, \vec y)
=  G_p(\vec x, \vec y) +G_{np}(\vec x, \vec y)\,.
$$
$G_{np}$ satisfies
$$
\nabla^2
G_{np}(\vec x, \vec y)
= {1 \over L_1L_2}\,.
$$
If one chooses $G_{np}$
(there is a gauge-like degree of freedom for $G_{np}$) as
\begin{equation}
G_{np}(\vec x, \vec y)
=  {({\rm Im }\, z)^2 \over 2 \, {\rm Im} \,\tau}
= {(x_2 - y_2)^2 \over 2 L_1 L_2}\,,
\label{Gnp}
\end{equation}
then
\begin{equation}
G(\vec x, \vec y)
= {1\over 4\pi} \ln \left| {\theta_1(z|\tau)
\over \theta_1{'}(0|\tau)} \right|^2\,.
\label{Green}
\end{equation}

The gauge generator of this hamiltonian system is
\begin{equation}
B \equiv i {\mu \over \hbar}
\int \! d \vec x \, b^{(1)}(x) \Lambda(x)\,,
\end{equation}
where $\Lambda(x)$ is an arbitrary real periodic function.
$a_i^{(1)}(x)$, $a_i(x)$ and
$\psi(x)$ are  gauge-dependent since
$$ [B, a_i^{(1)}(x)] = \partial_i \Lambda(x)\,,
\quad
[B, a_i(x)] = \partial_i \Lambda(x)\,,
$$
\begin{equation}
[B, \psi(x)] = {-ie\over \hbar c} \Lambda (x) \psi(x)
\end{equation}
whereas $\psi^{(1)}(x)$ is gauge invariant
\begin{equation}
[B, \psi^{(1)}(x)] =0\,.
\end{equation}
We emphasize that the readers should not confuse the gauge field
$ a_i^{(1)}(x)$ with $a_i^{({\rm eff})}(x)$.
$ a_i^{(1)}(x)$'s are satisfying the
commutation  relation given in Eq. (\ref{commfield}).
However, the commutation
relation between  $a_i^{({\rm eff})}(x)$'s is given as
\begin{equation}
[a_1^{({\rm eff})}(x), a_2^{({\rm eff})}(y)]
= i {\hbar \over \mu L_1L_2}\,,
\end{equation}
which is coordinate-independent.

Since $b^{(1)}(x)$ commutes
with any other operators in  ${\cal H}^{(1)}$,
and since we consider the gauge
invariant  operators and physical states,
we may drop  $b^{(1)}(x)$ completely.
Therefore, the local gauge fields  disappear
from the hamiltonian density,
\begin{equation}
{\cal H}^{(1)} (x)
=  {1 \over 2m} \left| (i \hbar \partial_i -
{e\over c} A_i^{[1]}(x) )\psi^{(1)}(x) \right|^2\,,
\label{hamil11}
\end{equation}
where
\begin{equation}
A_i^{[1]}(x)= {\theta_i \over L_i} -(Q + 1)
\tilde x_i -{e \over \mu c}
\epsilon_{ij} \partial_{x^j}
\int \! d \vec y G_p(\vec x, \vec y) J_0(y)\,.
\label{A1i}
\end{equation}

The effective hamiltonian density
and the newly  defined fields enable us to construct  the
physical state by simply applying operators consisting of
$\psi^{(1)+}$ and
$\theta_i$'s on the vacuum $|0>$
which satisfies  $b^{(1)}|0>=J_0|0>=0$
($\xi |0>=0$).
The presence of non-commuting zero-modes  $\theta_i$'s however,
prohibits  simple interpretation for the effective hamiltonian density.
Furthermore, the system can possess
a large gauge transformation invariance.
Suppose we transform
the  zero-modes by a constant $\Lambda_i$;
$$
\theta_i \rightarrow \theta_i +
\Lambda_i
$$
and the fermi-field by
$$ \psi^{(1)}(x)  \rightarrow
\exp (-i \phi_i(\vec x)) \psi^{(1)} (x)
$$
where
$\phi_i(\vec x)= {e \Lambda_i x^i \over \hbar c L_i}$
($i=1,2$, not sum on $i$).
This transformation leaves ${\cal H}^{(1)}$ invariant.
This transformation is, however,
not a gauge one since
$\phi_i(\vec x)$ does not satisfy the periodic boundary condition.
On the other hand, for the ``compact'' $U(1)$ gauge group
(equivalent to a circle) $\phi_i(\vec x)$ is
defined modulo $2\pi$.
In this case, this transformation becomes the large gauge
transformation if
${e \Lambda_i \over \hbar c L_i}= 2\pi n_i $,
where $n_i$ is an integer.
We, therefore, perform another
canonical transformations
and obtain an effective hamiltonian density which has
no explicit dependence of the zero-modes.
In this picture, the invariance under
the possible large gauge transformation is manifest.

The canonical transformation is explicitly given as
\begin{equation}
V_2(t)
= \exp \left[ {ie \theta_2^{(2)} \over \hbar c}
\int d \vec y J_0 ^{(2)}(y) {y^2 \over L_2}
\right]\,,
\label{V2}
\end{equation}
and the new fields are given as
\begin{eqnarray}
\psi^{(1)}(x) & =& V_2(t) \psi^{(2)}(x) V^+_2(t)
= \psi^{(2)}(x) \exp \{-i {e x^2 \theta_2^{(2)} \over \hbar c L_2}\}
\nonumber\\
\theta_1 (t)  &=&
V_2(t) \theta_1 ^{(2)}(t) V_2^+(t)
=\theta_1 ^{(2)}(t)
+ \phi \int d \vec y J_0 ^{(2)}(y) {y^2 \over L_2}
\nonumber\\
\theta_2 (t) &=& V_2(t)\theta_2^{(2)} (t)V_2^+(t)
=\theta_2^{(2)} (t)\,,
\label{field2def}
\end{eqnarray}
and $J_0^{(2)} (x)= J_0(x)$.
The hamiltonian density becomes
\begin{equation}
{\cal H}^{(2)}(x)
= {1 \over 2m}
\left| (i \hbar \partial_i
- {e\over c} A_i^{[2]}(x)) \psi^{(2)}(x) \right|^2\,,
\label{hamil2}
\end{equation}
where we have new effective gauge-like terms
\begin{eqnarray}
A_1^{[2]} (x) &= &{\theta_1^{(2)} \over L_1} -\tilde x_1 Q
+ \int \! d \vec y J_0(y) \tilde y_1 - \phi \partial_{x^2}
\int \! d \vec y G_p(\vec x, \vec y) J_0(y)
\nonumber \\
&=& {\theta_1^{(2)} \over L_1 } - \phi \partial_{x^2}
\int \! d \vec y G(\vec x, \vec y) J_0(y)\,,
\nonumber\\
A_2^{[2]} (x) &= & \phi \partial_{x^1}
\int \! d \vec y G(\vec x, \vec y) J_0(y)\,,
\label{A2i}
\end{eqnarray}
noting that $\partial_{x^1}G_{np}(\vec x, \vec y)=0$
from Eq. (\ref{Gnp}).

We remark that $\psi^{(2)}$ is invariant under  $\theta_2
\rightarrow  \theta_2 + 2\pi n_2 {\hbar c \over e}$
whereas $\psi^{(1)}$ is not:
$$ \psi^{(2)}(x) \rightarrow \psi^{(2)}(x)\,;\quad
\psi^{(1)}(x) \rightarrow  \exp \{-i 2\pi n_2{x^2 \over L_2} \}
\psi^{(1)}(x)\,. $$
Not only for that, $\theta_2$
disappears in ${\cal H}^{(2)}$
and the resulting effective gauge-like term
$A_i^{[2]}$ commutes each other.
With one more transformation using
\begin{equation}
V_3(t) = \exp \left[ {ie \theta_1^{(3)} \over \hbar c} \int d
\vec y J_0 ^{(3)}(y) {y^1 \over L_1} \right]\,,
\label{V3}
\end{equation}
we get
\begin{equation}
{\cal H}^{(3)} (x)
= {1 \over 2m} \left| (i \hbar \partial_i -
{e\over c} A_i^{[3]}(x)) \psi^{(3)}(x) \right|^2\,,
\label{hamil3}
\end{equation}
where
\begin{eqnarray}
\psi^{(2)}(x)
& =& V_3(t) \psi^{(3)}(x) V^+_3(t)
= \psi^{(3)}(x)
\exp \{-i {e x^1\theta_1^{(3)} \over \hbar c L_1}\}
\nonumber\\
\theta_1^{(2)} (t) &=& V_3(t) \theta_1 ^{(3)}(t) V_3^+(t)
=\theta_1 ^{(3)}(t)
\nonumber\\
\theta_2^{(2)} (t)
&=& V_3(t)\theta_2^{(3)}(t) V_3^+(t)
= \theta_2^{(3)}(t)  - \phi \int d \vec y J_0 (y) {y^1 \over L_1}
\nonumber\\
A_i^{[3]}(x) &= & - \phi
\epsilon_{ij} \partial_{x^j}
\int \! d \vec y G(\vec x, \vec y) J_0(y)\,,
\label{field3def}
\end{eqnarray}
and $J_0^{(3)} (x)= J_0(x)$.

This is the desired form of
the hamiltonian density we are looking for,
which amounts to the one
what one usually obtains by solving the Gauss constraint
if one neglects
the $x$-independent term in $A_i^{[3]}(x)$,
$\int d \vec y J_0 (y) \tilde y_i$ in
Eq.(\ref{field3def}). In this canonical formalism,
the effective gauge  fields,
$A_i^{[3]}(x)$ satisfy the Gauss constraint equation naturally,
$$
b^{({\rm eff})} (x)
\equiv -\epsilon_{ij} \partial_i A^{[3]}_j (x)
= - \phi J_0(x)\,.
$$
It is interesting to see that the commutation  relation  for
$\Pi_i^{[N]}(x)
\equiv i\hbar \partial_i - {e\over c} A_i^{[N]}(x)$
($N= 1,2,3$) is given as
\begin{equation}
[\Pi_1^{[N]}(x), \Pi_2^{[N]}(x)]
= -i\hbar {e\phi \over c}J_0(x) \,.
\end{equation}

The crucial point of our canonical formalism lies in
introducing  the (local and  large)
gauge-invariant (composite) matter field
$\psi^{(3)}(x)$.
$\psi^{(3)}(x)$ is related with the original matter field
$\psi(x)$ as
\begin{eqnarray}
\psi(x) &=& \exp -i(\eta + {e \over \hbar c} \int
d \vec y G_p (\vec x, \vec y)  \partial_{y^j} a_i^{(3)}(y))
\nonumber\\
&\times&
\exp ( {ie^2x^2 \over \mu\hbar c^2 L_2}
\int d \vec y J_0^{(3)} (y) {y^1 \over L_1} )
\nonumber\\
&\times&
\exp ( -i {ex^2 \theta_2^{(3)} \over \hbar c L_2})
\exp ( -i {ex^1 \theta_1^{(3)} \over \hbar c L_1})
\psi^{(3)} (x)\,,
\label{psi-psi3}
\end{eqnarray}
where $J_0^{(3)} (y) = J_0(y)$.
{}From this, we can deduce the gauge
transformation property of $\psi(x)$.
Since $a_i^{(3)} (x)$ is decoupled from
the hamiltonian  density ${\cal H}^{(3)} (x)$,
we can allow the gauge
transformation for $a_i^{(3)} (x)$
without transforming $\psi^{(3)} (x)$,
\begin{eqnarray}
a_i^{(3)} (x)
&\rightarrow& a_i^{(3)} (x) + \partial_i \Lambda (x)
\nonumber\\
\theta_i^{(3)} (t)
&\rightarrow& \theta_i^{(3)} (t)
+ 2 \pi n_i {\hbar c \over e}\,.
\end{eqnarray}
This induces the gauge transformation
for $\psi(x)$ and $a_i(x)$ as
\begin{eqnarray}
a_i(x)  &=& a_i^{(3)} (x)
- \phi \epsilon_{ij} \partial_{x^j} \int d\vec y J_0
(y) G(\vec x, \vec y) - \delta_{i2} {\phi \over L_2}
\int d\vec y J_0 (y) {y^1 \over L_1}
\nonumber\\
&\rightarrow&  a_i(x) + \partial_i \Lambda (x)\,,
\nonumber\\
\theta_i(t)  &=& \theta_i^{(3)} (t) + \phi \epsilon_{ij} \int d\vec
y J_0 (y) {y^j  \over L_j}
\nonumber\\
&\rightarrow& \theta_i(t) + 2\pi n_i {\hbar c \over e}\,,
\nonumber\\
\psi(x) &\rightarrow& \psi(x) \exp -i
\left[{e \Lambda (x) \over \hbar c}
+ 2\pi {n_1 x^1 \over L_1} +2\pi {n_2 x^2
\over L_2} \right]\,. \label{gauget}
\end{eqnarray}

The physical operators are products of $\psi^{(3)} (x)$'s and
$\psi^{(3)+}(x)$'s, and the vacuum $|0 > $
satisfies $b^{(3)} |0 > = J_0 |0 > = 0$.
In evaluating the expectation value of
the operators between the physical
states, we do not need the gauge sector of
the vacuum and therefore, degeneracy
structure resulting from the zero-modes $\theta_1$
and $\theta_2$ does not
contribute as far as the matter field $\psi^{(3)}(x)$ is concerned.

\section{Many-body quantum mechanics}

The $N$-particle wavefunction is defined as
\begin{equation}
\Phi (1, \cdots, N)
\equiv < 0|\psi^{(3)} (x^{(1)}) \cdots \psi^{(3)} (x^{(N)}) |N >
\label{Nwave}
\end{equation}
We assume that all the coordinates
of the particles lie in a fundamental domain.
$|0 >$ is the vacuum which satisfies $J_0|0 > = 0$
and $|N >$ is the $N$-body
Heisenberg state vector.
We suppress the vacuum degeneracy index due to the
gauge sector.  The wavefunction is gauge invariant
since
$\psi^{(3)}(x)$ is (local and large) gauge invariant.

The Schr\"odinger equation is given as
\begin{equation}
i\hbar {\partial\Phi \over \partial t} (1, \cdots, N)
= \sum_{p=1}^N < 0 | \psi^{(3)} (x^{(1)})
\cdots i\hbar {\partial\psi^{(3)} (x^{(p)})  \over \partial
t} \cdots \psi^{(3)} (x^{(N)}) |N > \,.
\end{equation}
To evaluate this, we need the Heisenberg equation
for the fermi-field $\psi^{(3)} (x)$,
$$
i\hbar {\partial\psi^{(3)} (x)
\over \partial t} = [\psi^{(3)} (x), H]$$
$$
= {1 \over 2m} \left[ \right.
(i\hbar\partial_{x^i} - {e \over e} A_i^{[3]} (x))^2 +
({e \over c})^2 \int d\vec y \psi^{(3)+} (y) K_i(y,x) K_i(y,x)\psi^{(3)} (y) $$
$$ - ({e \over c}) \int d\vec y \psi^{(3)+} (y) (i\hbar
\stackrel{\leftarrow}\partial_{y^i} -
({e \over c}) A_i^{[3]}  (y)) K_i(y,x)
\psi^{(3)} (y) $$
\begin{equation}
 - ({e \over c}) \int d\vec y \psi^{(3)+} (y))K_i(y,x)
 (i\hbar \vec \partial_{y^i} -
({e \over c}) A_i^{[3]}  (y)) \psi^{(3)} (y)
\left. \right] \psi^{(3)} (x) \,,
\label{NSch}
\end{equation}
where
\begin{equation}
K_i(x,y) = - \phi \epsilon_{ij} \partial_{x^j} G(\vec x, \vec y)\,.
\end{equation}
We put $\psi^{(3)+}$ and $A_i^{[3]}$
to the far left in each term such that the
operators vanish when they act on
the vacuum $<0 |$ and used the identity
$$
[\psi^{(3)} (x), A_i^{[3]} (y)]  = K_i (y,x) \psi^{(3)} (x)\,.$$

The  one-particle wavefunction satisfies the Schr\"odinger equation
\begin{equation}
i\hbar {\partial\Phi \over \partial t} (x)
= < 0 | i\hbar {\partial\psi^{(3)}
(x) \over \partial t} |1 >
= - {\hbar^2  \over 2m} \nabla ^2 \, \Phi(x)\,.
\end{equation}
The equation does reduce to the free one.
For the two-body case, we have
\begin{eqnarray}
i\hbar {\partial\Phi \over \partial t} (1,2)
&=& < 0 | \psi^{(3)} (x^{(1)})
i\hbar {\partial\psi^{(3)} (x^{(2)}) \over \partial t} +
i\hbar {\partial\psi^{(3)} (x^{(1)})
\over \partial t} \psi^{(3)} (x^{(2)}) |2 >
\nonumber\\
&=& {1 \over 2m} \{(i\hbar \partial_i^{(1)}
- {e  \over c} {\cal A}_i (1,2))^2 +
(i\hbar \partial_i^{(2)} - {e  \over c}
{\cal A}_i (2,1))^2 \} \Phi(1,2)\,,
\end{eqnarray}
where
\begin{equation}
{\cal A}_i (p,r) =  - \phi \epsilon_{ik} \partial_k^{(p)} G(p,r)\,.
\end{equation}
In general for the $N$-particle case, we have
\begin{equation}
i\hbar {\partial\Phi \over \partial t} (1, \cdots, N)
= {1 \over 2m} \sum^N_{p=1} \{i\hbar \partial_i^{(p)} - {e  \over c}
\sum^N_{r=1(\ne p)} {\cal A}_i (p,r) \}^2
\Phi(1, \cdots, N)\,. \label{NSchp}
\end{equation}

Finally, the Aharonov-Bohm gauge potential,
${\cal A}_i (\vec x, \vec y)$ can be
transformed away through the singular gauge transformation
as in the infinite plane case.
Explicitly, using the Green's function
in Eq. (\ref {Green}) we have
$$ {\cal A}_i (p,r)
= - \partial_i^{(p)}
\left[ {i\phi \over 4\pi}
\ln {\theta (z_{pr} |  \tau) \over \theta^* (z_{pr} |  \tau) }
\right]\,,$$
if we neglect the singular parts at the coincident points.
As the result, the Schr\"odinger equation becomes
the free one, whereas  the transformed
wave-function (anyonic wavefunction) is multivalued
$$
\Phi^{({\rm anyon})} (1, \cdots, N) = \prod^N_{p>r}
\left({\theta (z_{pr} |  \tau)
\over \theta^* (z_{pr} |  \tau) } \right)^
{\nu \over 2} \Phi (1, \cdots, N)\,,$$
where $\nu = e^2 / (2\pi\hbar\mu c^2)$
and $\phi=\nu\phi_0$ with the unit flux
quantum $\phi_0 = hc/e$.

\section{Periodic property}

To describe a closed surface such as a torus,
one usually adopts the covering
space, which consists of the repeated domains
of the fundamental one, and
identifies the edges of the fundamental domain with each other.
Then the non-contractible loop on the closed surface
is identified with a line on the
fundamental domain from one edge to the other.

We have considered so far the hamiltonian system
on the fundamental domain,
$D_{(0,0)}$ $(0 \le x^1 < L_1, 0 \le x^2 < L_2)$.
To describe the hamiltonian density
outside the fundamental domain,
let us introduce the domain $D_{(m,n)}$ for
definiteness, $(x{'}^1 = x^1 + mL_1, x{'}^2 = x^2 + nL_2)$
where $x^i$ lies on
$D_{(0,0)}$ and $m$,  $n$ are integers.
(In the following, we reserve the
unprimed coordinates for the ones in the fundamental domain
$D_{(0,0)}$ and primed for the domain $D_{(m,n)}$).

Obviously, the effective hamiltonian density
on $D_{(m,n)}$ can be written as
\begin{equation}
{\cal H}^{(3)} (x{'})
= {1 \over 2m} \left|
(i \hbar \partial_i - {e\over c} A_i^{[3]} (x{'})) \psi^{(3)} (x{'})
\right|^2\,,
\label{hamil3prime}
\end{equation}
where
$$A_i^{[3]} (x{'})
= - \phi \epsilon_{ij} \partial_{x^j}
\int \! d \vec y G(\vec x{'}, \vec y\,{'}) J_0(y{'})\,.$$
Noting the relation,
$G(\vec x, \vec y) = G(\vec x{'}, \vec y\,{'})$,
we have
\begin{equation}
A_i^{[3]} (x{'})
= - \phi \epsilon_{ij} \partial_{x^j}
\int \! d \vec y G(\vec x, \vec y) J_0(y{'})\,.
\label{field3defprime}
\end{equation}

This hamiltonian density on $D_{(m,n)}$ is canonically
related with the one on
$D_{(0,0)}$
\begin{equation}
{\cal H}^{(3)} (x{'})
= T_{(m,n)} {\cal H}^{(3)} (x)  T^+_{(m,n)}\,,
\end{equation}
where
\begin{eqnarray}
J_0 (x{'}) &=& T_{(m,n)} J_0 (x) T^+_{(m,n)}\,,
\nonumber\\
b^{(3)}(x{'}) &=& T_{(m,n)} b^{(3)} (x) T^+_{(m,n)}\,,
\nonumber\\
\psi^{(3)}(x{'})
&=& T_{(m,n)} \psi^{(3)} (x) T^+_{(m,n)}\,. \label{fieldprime}
\end{eqnarray}
The explicit translation operator
along the non-contractible loop is given as
\begin{equation}
T_{(m,n)} =  T_1^m T_2^n
\end{equation}
where
\begin{eqnarray}
T_1 &\equiv & \exp -L_1 \int d\vec y \{\psi^{(3)+} (y)
\partial_{y^1} \psi^{(3)} (y)
+ i {\mu \over \hbar} a_1^{(3)} (y) \partial_{y^1} a_2^{(3)} (y) \}
\nonumber\\
T_2 &\equiv & \exp -L_2 \int d\vec y
\{\psi^{(3)+} (y) \partial_{y^2} \psi^{(3)} (y)
+ i {\mu \over \hbar} a_1^{(3)} (y) \partial_{y^2} a_2^{(3)} (y) \}\,.
\end{eqnarray}
We also note that the translation operators
connect the gauge field as
\begin{equation}
a_i^{(3)}(x{'}) = T_{(m,n)} a_i^{(3)} (x) T^+_{(m,n)}\,.
\end{equation}
The operators $T_1$ and $T_2$ commute each other,
\begin{equation}
T_1T_2  = T_2T_1\,, \label{Tcomm}
\end{equation}
without the flux quantization
in contrast with the previous analyses.
The difference originates from the way
to treat the Gauss constraint.
We introduced the mean field mode $\xi$ in
$a_i(x)$ to yield a simple periodic condition
(see Eq. (\ref {fieldperiod})
below) whereas the previous analyses
require the quasi-periodic condition on the
gauge field $a_i(x)$ and the matter field $\psi(x)$, which results in
non-commuting translation operators
unless the flux is quantized.

On the torus, one can leave
the hamiltonian density
invariant under the translation along
the non-contractible loops as given in Eq. (\ref{fieldprime})
if one defines the matter field
as $\psi^{(3)} (x{'}) =
\psi^{(3)} (x) \exp iC_{mn},$
where $C_{mn}$ is a constant on $D_{(m,n)}$.  We can
choose the constant $C_{mn} = 0$ without losing any generality.
\begin{equation}
\psi^{(3)} (x{'}) = \psi^{(3)} (x) \,.
\label{psiperiod}
\end{equation}
This definition leads to the periodic condition
on $J_0$ and $b^{(3)}$ as
\begin{equation}
J_0 (x{'}) = J_0 (x) ; \quad
b^{(3)} (x{'}) = b^{(3)} (x)\,. \label{J0period}
\end{equation}
This makes the effective gauge field
$A_i^{[3]} (x{'})$ manifestly
translation-invariant,
\begin{equation}
A_i^{[3]} (x{'}) = A_i^{[3]} (x)\,, \label{Aperiod}
\end{equation}
and the translation-invariance of the hamiltonian density follows :
\begin{equation}
[T_i, {\cal H}(x)] = 0\,.  \label{THcomm}
\end{equation}
We note that the vacuum defined
on the fundamental domain remains the same on
the covering space.
Therefore, the physical state is the same on all the
covering space and the Gauss contraint is identically satisfied.

The equal-time commutation relation
of $J_0(x)$ with $\psi^{(3)} (y{'})$ is
given as
\begin{equation}
[J_0(x), \psi^{(3)} (y') ] = [J_0(x), \psi^{(3)} (y) ]
= \psi^{(3)} (y) \delta (\vec x - \vec y)\,.
\end{equation}
One can also define the periodic property
of the gauge field $a_i^{(3)}$ up to a
gauge transformation,
which leaves the field strength $b^{(3)}$ invariant as
given in Eq.(\ref {J0period}),
\begin{equation}
a_i^{(3)} (x')
= a_i^{(3)} (x) + \partial_i \Omega (x)\,,  \label {aiperiod}
\end{equation}
where $\Omega(x)$ is an arbitrary periodic function.

The original hamiltonian density
${\cal H}(x')$ on $D_{(m,n)}$ can be obtained
from the effective one ${\cal H}^{(3)}(x')$
if one uses the canonical operators,
\begin{eqnarray}
V_{(m,n)\,1} (t)
&=& \exp i \left[ Q \eta^{(1)}
+ {e\over \hbar c} \int\! \int
d\vec x d \vec y J_0 (x{'} ) G_p (\vec x, \vec y)
\partial_k a_k^{(1)}(y{'} ) \right] \,, \nonumber\\
V_{(m,n)\,2} (t)
&=&  \exp \left[ {ie \theta_2^{(2)} \over \hbar c} \int d \vec
y J_0 (y{'}) {y^2 \over L_2} \right]\,, \nonumber\\
V_{(m,n)\,3} (t)
&=& \exp \left[ {ie \theta_1^{(3)} \over \hbar c} \int d \vec y
J_0 (y{'}) {y^1 \over L_1} \right] \label{Vprime}\,.
\end{eqnarray}
where the integration
$\int d\vec x$ denotes for $\int_{0\le x^1<L_1, 0\le x^2
<L_2} d \vec x$.
The operators are the same as the ones given in Eqs. (\ref
{V1}, \ref {V2},  \ref {V3})
except that the field operators are replaced by the
ones on $D_{(m,n)}$.
The hamiltonian density becomes
\begin{equation}
{\cal H} (x{'}) = {1 \over 2m}
\left| (i \hbar \partial_i - {e\over c} a_i (x{'}))
\psi (x{'}) \right|^2\,, \label{hamilprime}
\end{equation}
where
\begin{eqnarray}
\psi(x{'})
&=& \exp -i(\eta + {e \over \hbar c}
\int d \vec y G_p (\vec x, \vec y)
\partial_{y^j} a_i^{(3)}(y{'})) \nonumber\\
&\times& \exp ( {ie^2x^2 \over \mu\hbar c^2 L_2}
\int d \vec y J_0^{(3)} (y) {y^1 \over L_1} )
\exp ( -i {ex^2 \theta_2^{(3)} \over \hbar c L_2})
\exp ( -i {ex^1 \theta_1^{(3)} \over \hbar c L_1})   \psi^{(3)} (x) \,,
\nonumber\\
a_i(x{'})
&=& a_i^{(3)} (x{'})
- \phi \epsilon_{ij} \partial_{x^j}
\int d\vec y G(\vec x, \vec y) J_0 (y)
- \delta_{i2} {\phi \over L_2}
\int d\vec y J_0 (y) {y^1 \over L_1} \,.
\label{psiprime}
\end{eqnarray}
The periodic property of the fields are easily deduced using Eqs. (\ref
{psiperiod}, \ref {aiperiod}),
\begin{eqnarray}
\psi (x{'}) &=& \psi(x) \exp - i {e \Omega (x) \over \hbar c}\,,
\nonumber\\
a_i (x') &=& a_i (x) + \partial_i \Omega (x)\,,  \label {fieldperiod}
\end{eqnarray}
which demonstrates that the periodic property
for $\psi(x)$ and $a_i(x)$ is
given as a local gauge transformation.
The commutation relations of the fields
are given as
$$ \{\psi(x{'}), \psi(y{''})\} = 0 $$
$$ \{\psi(x{'}), \psi^+(y{''})\} = \delta (\vec x - \vec y)$$
$$[a_1 (x{'}), a_2(y{''}) ]
= i {\hbar \over \mu}\delta (\vec x - \vec y)\,,$$
where $x'$ and $y''$ may lie in different domains.
We leave the explicit
construction of the two commuting translation operators
$VT_i V^+ (i=1,2)$ for
the original picture in terms of the original fields
to the devoted reader.

We emphasize that the canonical operators
in Eq. (\ref {Vprime}) are not the unique choice.
One may adopt a different choice
for the canonical operators as,
\begin{eqnarray}
V^{'}_{(m,n)\,1} (t)
&=& \exp i \left[ Q \eta^{(1)} + {e\over \hbar c}
\int\! \int d\vec x d \vec y J_0 (x{'} )
G_p (\vec x{'}, \vec y{'}) \partial_k
a_k^{(1)}(y{'}) \right] \,, \nonumber\\
V^{'}_{(m,n)\,2} (t)
&=&  \exp \left[ {ie \theta_2^{(2)} \over \hbar c}
\int d \vec y J_0 (y{'}) {y{'}^2 \over L_2} \right]\,, \nonumber\\
V^{'}_{(m,n)\,3} (t)
&=& \exp \left[ {ie \theta_1^{(3)} \over \hbar c}
\int d \vec y J_0 (y{'}) {y{'}^1 \over L_1}
\right]\,. \label{Vprime1}
\end{eqnarray}
Here, the difference lies in that the coordinates are also primed.
In this case,
the hamiltonian density becomes the same form
${\cal H} (x{'})$ in Eq. (\ref {hamilprime}).
The difference comes to the periodic property of the matter
field $\psi{'} (x{'})$
(we put the prime to distinguish this from $\psi (x{'})$
in Eq. (\ref {psiprime})) since the matter field is
given as
\begin{eqnarray}
\psi{'}(x{'})
&=& \exp -i(\eta + {e \over \hbar c}
\int d \vec y G_p (\vec x, \vec y)
\partial_{y^j} a_i^{(3)}(y{'}))
\nonumber\\
&\times&
\exp ( {ie^2x{'}^2 \over \mu\hbar c^2 L_2}
\int d \vec y J_0^{(3)} (y) {y{'}^1 \over L_1} )
\exp ( -i {ex{'}^2 \theta_2^{(3)} \over \hbar c L_2})
\exp ( -i {ex{'}^1 \theta_1^{(3)} \over \hbar c L_1})   \psi^{(3)} (x)
\,.\label{psiprime1}
\end{eqnarray}
It shows that the periodic property is as complicated
as can be and it has not
only the number operator dependent phase
but also the $\theta_i$ dependent one
$\exp \{-i {e\theta_i \over \hbar c} \}$
in addition to the phase factor
$\exp \{-i {e\Omega (x) \over \hbar c} \}$
in Eq. (\ref {fieldperiod}).
However, the local and large gauge transformation property
of the two matter
fields in Eqs.(\ref {psiprime}, \ref {psiprime1})
is exactly the same.

Let us consider the many anyon wavefunctions
whose coordinates are located in different domains.
Similarly in Eq. (\ref {Nwave}),
we define the wavefunction as
\begin{equation}
\Phi (\{1\}, \cdots, \{N\})
\equiv < 0|\psi^{(3)} (\{1\}) \cdots \psi^{(3)} (\{N\}) |N >
\label{Nwave1}
\end{equation}
where the coordinates
$\{i\} \equiv (x^{(i)1} + m^{(i)} L_1, x^{(i)2} + n^{(i)} L_2)$
with $m^{(i)}$ and $n^{(i)}$ integers
may lie in any mixed domain.
The Schr\"odinger equation is the same
as the one given in Eq. (\ref {NSchp}) since
the hamiltonian density ${\cal H}^{(3)} (x{'})$
is translation-invariant :
\begin{equation}
i\hbar {\partial\Phi \over \partial t}
(\{1\}, \cdots, \{N\})
= {1 \over 2m} \sum^N_{p=1}
\{i\hbar \partial_i^{(p)} - {e  \over c}
\sum^N_{r=1(\ne p)}
{\cal A}_i (p,r)\}^2 \Phi(\{1\}, \cdots, \{N\}) \,.
\label{NSchp1}
\end{equation}
The hamiltonian has the periodic gauge-like potential
since it contains the
coordinate function of $x^{(i)} $ rather than $x^{(i)}{'} $.
One should note that the
wavefunction is trivially periodic
\begin{equation}
\Phi (\{1\}, \cdots, \{N\}) =  \Phi (1, \cdots, N)\,,
\end{equation}
and it maintains the fermionic exchange property
\begin{equation}
\Phi (\cdots, \{i\}, \cdots, \{j\}, \cdots)
=  - \Phi (\cdots, \{j\}, \cdots, \{i\}, \cdots)\,,
\end{equation}
due to the periodic property of the matter field
in Eq. (\ref {psiperiod}).

\section{Summary and Discussion}

We have analyzed the Chern-Simons theory
coupled to non-relativistic matter
field on a torus.
Quantizing the field first and performing canonical
transformation we obtain
an effective theory in terms of matter field
$(\psi^{(3)} (x))$ only
such that it manifests the gauge invariance and the
translation invariance along the non-contractible loops.
The periodic property
of the new fields are explicitly obtained.
This assigns the  periodic property of the
original fields through the canonical transformation.
It is noted, however,
the periodic property of the original fields
may not be uniquely determined, which depends
on the choice of the canonical transformation.
{}From the effective hamiltonian we construct
the Schr\"odinger equation for many anyons.
It has the periodic Aharonov-Bohm potential
of the singular gauge form adapted to the boundary
condition and when transformed away
it reduces to the free equation with
multi-valued (anyon) wavefunction.

Our analysis does not need the flux quantization.
In the literature, one attempts to solve
the Gauss constraint on the $N$-particle state,
$\int d\vec y (b(y) + \phi J_0 (y)) |N > = 0$,
and requires
$\nu (\equiv {e^2 \over 2\pi \hbar \mu c^2})$
be a rational value since
$\int d\vec y b(x) / \phi_0$ $(\phi_0$
is the unit flux quantum)
and the number operator $Q$ are assumed
to be independent integers.
In our analysis, however,
the two eigenvalues are related each other,
 $\int d\vec y b(y) |N > =  - \phi N|N > $
 due to the mean field mode $\xi$ in the gauge field and
$Q|N > = N|N > $
since
$$ |N > \sim \int d \vec x^{(1)}
\cdots  d \vec x^{(N)} \psi^{(3)+} (1) \cdots
\psi^{(3)+} (N) | 0 >\,, $$
and the Gauss law is automatically satisfied
as mentioned following Eq. (\ref {comment}).
In addition, the translation operators
we constructed on the torus commute each other.
The effective gauge-like potential can be defined as a
periodic one from the beginning
(See Eq. (\ref {Aperiod})) and therefore, the
translation operators automatically
commute each other as given
in Eqs. (\ref {Tcomm}, \ref {THcomm}) for any value of $\nu$.

We also demonstrate that
the local and large gauge
transformation property
for the matter field does not fix the periodic condition
for the matter field in section V{}.
In addition, the periodic property of the gauge
invariant wavefunction
may depend on the choice of its definition.
Suppose one assigns the periodic property of $\psi(x)$
as given in Eq. (\ref {fieldperiod})
and choose the local
and large gauge invariant wavefunction as
$$\Phi_h (\{1\}, \cdots, \{N\}) $$
$$
\equiv < 0|
\exp (i \sum^N_{p=1} {e\{x^{(p)} \}^{ 1} \theta_1 \over \hbar c L_1} )
\exp(i \sum^N_{p=1} {e\{x^{(p)} \}^{ 2} \theta_2 \over \hbar c L_2} )
\psi^{(1)} (\{1\}) \cdots \psi^{(1)} (\{N\}) |N >\,,
$$
where $\psi^{(1)} (\{i\})$ is locally gauge invariant.
We can rewrite $\Phi_h$
as
$$\Phi_h (\{1\}, \cdots, (\{N\}) $$
$$
=  C_{(\{m\},\{n\})} (1, \cdots, N) < 0|
W_{( \tilde m , \tilde n )}(\theta_1, \theta_2)
\psi^{(3)} (\{1\}) \cdots \psi^{(3)} (\{N\}) |N >\,,
$$
where $C_{(\{m\},\{n\})} (1, \cdots, N)$ is a phase factor.
$W_{( \tilde m , \tilde n  )}(\theta_1, \theta_2)$
is the Wilson loop operator defined as
$$
W_{( \tilde m , \tilde n )} (\theta_1, \theta_2)
\equiv \exp (i \tilde m {e \theta_1 \over \hbar c })
\exp (i \tilde n {e \theta_2 \over \hbar c })\,,$$
with $ \tilde m  = \sum^N_{i=1} m^{(i)}$
and $ \tilde n= \sum^N_{i=1} n^{(i)}$.
The Schr\"odinger equation for $\Phi_h$
looks the same as that of $\Phi$ in Eq. (\ref {NSchp1}).
However, the Wilson loop will
detect the vacuum degeneracy due to
the zero-modes of the gauge fields
independently of the matter sector.
Therefore, the periodic condition
can be non-trivial as given in Ref.\cite{hosotani} due to
the Wilson loop and the wavefunction
can behave as the multi-component one
whose component is finite when $\nu$
(related with the Chern-Simons coefficient $\mu$) is rational.
It is not clear yet to us
which choice of the definition of the wavefunction is
physically relevant.

\acknowledgments

This work is supported in part by KOSEF No. 931-0200-030-2
and by SRC program through Seoul National University.


\begin {thebibliography}{99}

\bibitem {deser}  S. Deser, R. Jackiw and S. Templeton,
Ann. Phys. (N. Y.)  {\bf 140}, 372 (1982).
\bibitem {hagen}  C. Hagen, Ann. Phys. (N. Y.)
{\bf 157}, 342 (1984) ;
Phys. Rev. D {\bf 31}, 2135 (1985).
\bibitem {leinaas}  J. Leinaas and J. Myrheim,
Nuovo Cim. B {\bf 37}, 1 (1977) ;
G. Goldin, R. Menikoff and D. H. Sharp,
J. Math. Phys. {\bf 22}, 1664 (1981) ;
F. Wilczeck, Phys. Rev. Lett. {\bf 49}, 957 (1982) ;
Y. S. Wu, Phys. Rev. Lett. {\bf 52}, 2103 (1984).
\bibitem {goldhaber}  A. S. Goldhaber, R. Mackenzie and F. Wilczek,
Mod. Phys. Lett. A {\bf 4}, 21 (1989) ;
T. H. Hansson, M. Ro\u{c}ek, I. Zahed and S. C. Zhang,
Phys. Lett. {\bf B214}, 475 (1988).
\bibitem {semenoff}  G. W. Semenoff,
Phys. Rev. Lett. {\bf 61}. 517 (1988) ;
G. W. Semenoff, P. Sodano and Y. S. Wu,
{\it ibid}. {\bf 62}, 715 (1989) ;
G. W. Semenoff and P. Sodano,
Nucl. Phys. B {\bf 328}, 753 (1989).
\bibitem {matsuyama}  T. Matsuyama,
Phys. Lett. {\bf B228}, 99 (1989) ;
Phys. Rev. D {\bf 15}, 3469 (1990) ;
D. Boyanovsky, E. Newman and C. Rovelli, Phys.
Rev. D {\bf 45}, 1210 (1992) ;
R. Banerjee, Phys. Rev. Lett. {\bf 69}, 17 (1992) ;
Nucl. Phys. B {\bf 390}, 681 (1993).
\bibitem {swanson}   M. S. Swanson,
Phys. Rev. D {\bf 42}, 552 (1990).
\bibitem {cho}  K. H. Cho and C. Rim,
Int. J. Mod. Phys. A {\bf 7}, 381 (1992).
\bibitem {hong}  J. Hong, Y. Kim, P. Y. Pac,
Phys. Rev. Lett. {\bf 64}, 2230
(1990) ; R. Jackiw and E. J.  Weinberg,
{\it ibid}. {\bf 64}, 2234 (1990) ;
R. Jackiw, K. Lee and E. J. Weinberg,
Phys. Rev. D {\bf 42}, 3488 (1990) ;
C. Lee, K. Lee and H. Min,
Phys. Lett. {\bf B 252}, 79 (1990).
\bibitem {kim}  S. K. Kim and H. Min,
Phys. Lett. {\bf B281}, 81 (1992) ;
L. Hua and C. Chou, {\it ibid}. {\bf B308}, 286 (1993).
\bibitem {girvin}   For a review and references,
see S. M. Girvin and R. Prange,
{\it The Quantum Hall Effect\/} (Springer, New York, 1990)
and F. Wilczek, {\it Fractional Statistics and Anyon Superconductivity\/}
(World Scientific, Singapore, 1990).
\bibitem {wu}  Y. S. Wu, Phys. Rev. Lett. {\bf 53}, 111 (1984) ;
A. Polychronakos, Phys. Lett. {\bf B264}. 362 (1991) ;
C. Chou, Phys. Lett. {\bf A155}, 245 (1991) ;
K. H. Cho and C. Rim, Ann. Phys. {\bf 213}, 295 (1992) ;
G. V. Dunne, A. Lerda, S. Sciuto and C. A. Trugenberger,
Nucl. Phys. B {\bf 370}, 601 (1992) ;
K. H. Cho, C. Rim and D. S. Soh,
Phys. Lett. {\bf A164}, 65 (1992)
; Int. J. Mod. Phys. A {\bf 8}, 5101 (1993).
\bibitem {amelino}   G. Amelino-Camelia and L. Hua,
Phys. Rev. Lett. {\bf 69},
2875 (1992) ; K. H. Cho and C. Rim,
{\it ibid}. {\bf 69}, 2877 (1992).
\bibitem {sporre}  M. Sporre, J. J. M. Verbaarschot and I. Zahed,
Phys. Rev. Lett. {\bf 67}, 1813 (1991) ;
M. V. N. Murthy, J. Law, M. Brack and R. K. Bhaduri,
{\it ibid}. {\bf 67}, 1817 (1991).
\bibitem {polychro}  A. P. Polychronakos,
Ann. Phys. {\bf 203}, 231 (1990).
\bibitem {randjbar}  S. Randjbar-Daemi, A. Salam and J. Strathdee,
Phys. Lett. {\bf B240}, 121 (1990).
\bibitem {iengo}  R. Iengo and K. Lechner,
Nucl. Phys. B {\bf 346}, 551 (1990),
{\it ibid}. {\bf 346}, 551 (1991).
\bibitem {hosotani}  Y. Hosotani, Phys. Rev. Lett.
{\bf 62}, 2785 (1989) ; C.
-L. Ho and Y. Hosotani, Int. J. Mod. Phys. A {\bf 7}, 5797 (1992) ;
Phys. Rev. Lett. 70, 1360 (1993).
\bibitem {fayya}  A. Fayyazuddin,
Nucl. Phys. B {\bf 401}, 644 (1993).
\bibitem{ein} T. Einarsson, Phys. Rev. Lett. {\bf 64}, 1995 (1990).
\bibitem {jackiw}  R. Jackiw and S. -Y. Pi,
Phys. Rev. D {\bf 42}, 3500 (1990)
\bibitem {kimlee}  C. Kim, C. Lee, P. Ko, B. -H. Lee
and H. Min, Phys. Rev. D
{\bf 48}, 1821 (1993).
\bibitem {hagsu} C. R. Hagen and E. C. G. Sudarshan,
Phys. Rev. Lett. {\bf 64},
1690 (1990) ; Y. Hosotani,
{\it ibid}. {\bf 64}, 1691 (1990).

\end {thebibliography}
\end{document}